\documentclass[
aps,
prb,
reprint,
superscriptaddress,
showpacs,
footinbib
]{revtex4-1}

\usepackage[utf8x]{inputenc}
\usepackage{graphicx}
\usepackage{amsmath}
\usepackage[english]{babel}
\usepackage{siunitx}
\usepackage{todonotes}
\usepackage{nicefrac}
\usepackage{calc}
\usepackage{amssymb}

\usepackage{hyperref}
\hypersetup{colorlinks=false,urlcolor=blue,linkcolor=blue,citecolor=blue,pdftitle={TaSeDoping},pdfcreator={LaTeX2e,TeX Live 2012}}

\newcommand{\2}{$2H$-TaSe$_2$}
\newcommand{\name}[1]{{\scshape{#1}}}

\renewcommand{\eqref}[1]{Eqn.\,\ref{#1}}
\newcommand{\figref}[1]{Fig.\,\ref{#1}}

\newcommand{\tmdcs}{transition-metal di\-chalco\-genides}
\newcommand{\eels}{electron energy-loss spectroscopy}
\renewcommand{\epsilon}{\varepsilon}

\begin{document}

\title{Doping Dependence of the Plasmon Dispersion in $2H$-TaSe$_2$}

\date{\today}

\author{A.\,\surname{König}}
\email[email: ]{a.koenig@ifw-dresden.de}
\author{R.\,\surname{Schuster}}
\author{M.\,\surname{Knupfer}}
\author{B.\,\surname{Büchner}}
\affiliation{IFW Dresden, Institute for Solid State Research, P.\,O. Box 270116, D-01171 Dresden, Germany}
\author{H.\,\surname{Berger}}
\affiliation{Institut de Physique de la Mati\`ere Complexe, EPFL, CH-1015 Lausanne, Switzerland} 

\begin{abstract}
Inelastic electron scattering is applied to investigate the impact of potassium intercalation on the charge-carrier plasmon energy and dispersion in the charge-density wave (CDW) bearing compound \2. We observe an unususal doping dependence of the plasmon dispersion, which even changes sign on alkali addition.

In contrast to the continous energy shift of the plasmon position on doping at lowest momentum transfer, its dispersion changes in a rather discontinuous manner. We argue that the observed dynamics can be explained only in a picture, where complex phenomena are taken into account including the suppression of the CDW on doping as well as the interplay of the CDW and the plasma resonance.
\end{abstract}


\maketitle

\section{Introduction}
\label{sec:intro}

For decades the layered \tmdcs\ (TMDCs) have attracted considerable interest of researchers in different fields of physics and materials science. Besides the various anisotropic properties,~\cite{Brown1965, Quinn1966, vanMaaren1967, Wilson1969} the layered structure also exhibits a charge-density wave (CDW) phase transition within the $ab$~plane of the crystal.~\cite{Wilson1975, Moncton1975, Moncton1977} For \2 a detailed study of this phase transition---exhibiting an incommensurate superstructure followed by a lock-in transition to a commensurate ordering vector---has been, for example, proven by neutron scattering~\cite{Moncton1977} or x-ray diffraction experiments.~\cite{Leininger2011} The associated opening of a gap at the Fermi level was shown by  angle-resolved photoemission measurements.~\cite{Borisenko2008, Inosov2009} Reflectivity as well as resistivity investigations, however, show a partial appearance of a gap at the \name{Fermi} level already for temperatures far above the phase transition.%
~\cite{Vescoli1998}

In regard to the collective excitation of the conduction electrons, the TMDC materials revealed a further intriguing feature. The momentum dependence of the collective charge-carrier excitation---termed the \emph{plasmon}---was found to reveal a shift to lower energies with increasing momentum transfer, which contradicts the conventional behavior of a metal.~\cite{Schuster2009} Different investigations have made attempts to clarify this exceptional dispersion. In a semiclassical Ginzburg-Landau approach this effect was assigned to an interplay of the charge density fluctuations and the plasma resonance.~\cite{vanWezel2011a} Alternatively, recent band structure calculations attribute the negative dispersion to the shape of the bands crossing the \name{Fermi} level.~\cite{Cudazzo2012, Faraggi2012} However, the calculated loss-spectra exhibit a double peak structure around \SI{1}{eV}, which is in contrast to the experimentally obtained single peak forming the plasmon. In addition, it has to be mentioned that the plasmon width on doping becomes smaller by shifting to low energies,~\cite{Koenig2012} a fact that normally can be attributed to decreasing interband damping effects.~\cite{Paasch1970, Sturm1976}

The crystal structure of the TMDCs exhibits a large \name{van\,der\,Waals}\,gap between their hexagonal sandwich like building blocks stacked in the crystallographic $c$-axis. This gap is thought to be the typical place for all kinds of intercalates to arrange within the crystal.~\cite{Friend1987,Rouxel1979} On alkali metal intercalation (Na with Na/Ta$ = 0.3$ and Cs with Cs/Ta varying between 0.3 and~0.6) the band structure of  \2~\cite{Brauer2001, Pettenkofer1992} and related compounds, such as $1T$-Rb$_x$TaSe$_2$~\cite{Stoltz2003} and $1T$-Cs$_x$TaSe$_2$ (up to $x=0.5$),~\cite{Crawack2000} is reported to change notably.

\begin{figure*}[t]
  \centering
  \includegraphics{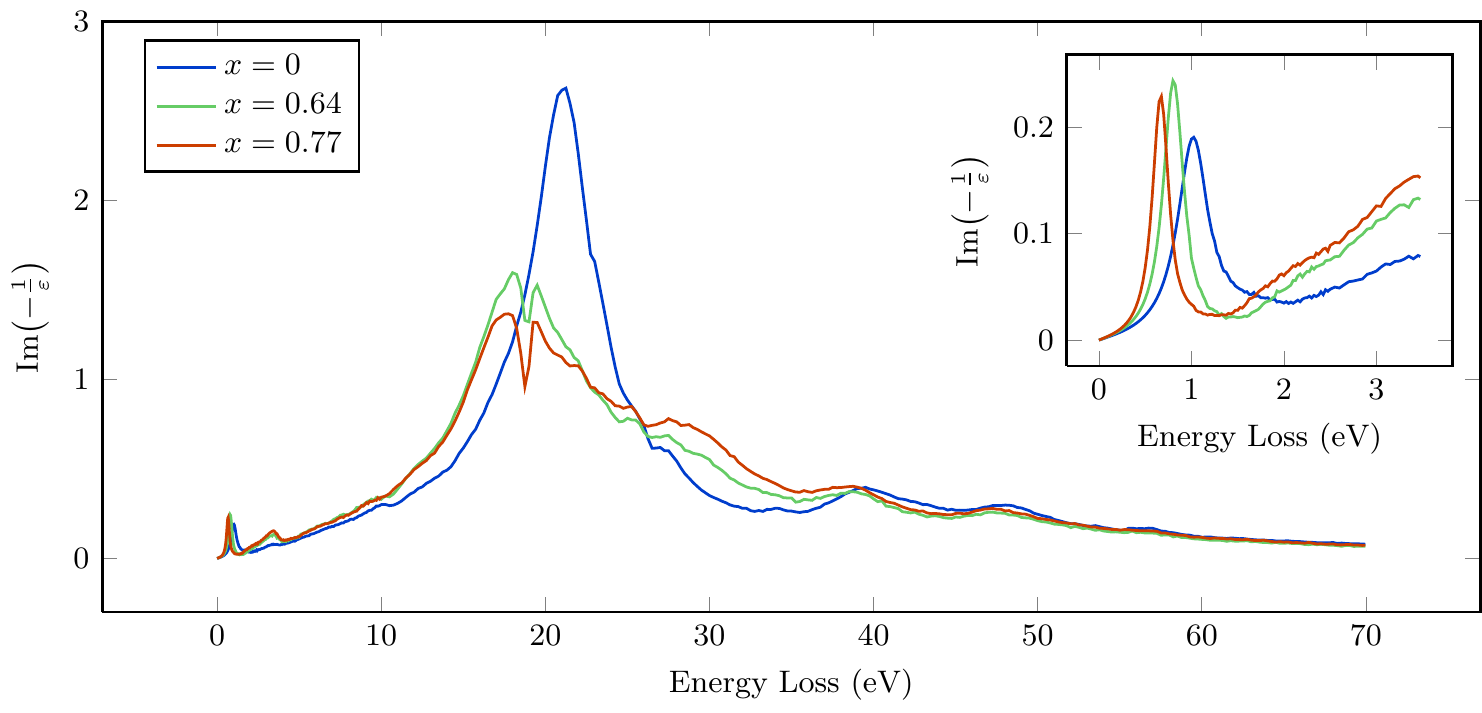}
  \caption{Loss spectra for different doping levels of $2H$-K$_x$TaSe$_2$ normalized via the KKA integration. The inset shows a detailed view of the charge carrier plasmons behavior on doping.\cite{Koenig2012}}
  \label{fig:kka_spectra}
\end{figure*}

Moreover, changes in structure can be observed in various polytypes of the TMDCs. Because of the sandwich configuration, in particular the $c$~axis is widened, while the spacing within the basal planes remains nearly unchanged.~\cite{Starnberg2000, Omloo1970} The latter is indicated by slight changes of the superstructure for different compounds such as the aforementioned $1T$-Cs$_x$TaSe$_2$~\cite{Crawack2000} as well as for K$_x$TiS$_2$.~\cite{Pronin2001} CDW superstructures can be ruled out because of a suppression of the ordered state on doping, as, for example, in $2H$-K$_x$TaS$_2$.~\cite{Biberacher1985}

In the case of silver-doped TMDCs, further structural changes are discussed in a staging framework, which is in close analogy to intercalated graphite compounds. New scattering channels are introduced by a slight disarrangement of the chalcogenide atoms by the added silver atoms. The stages then can be observed at doping rates of one- and two-thirds, respectively, for Ag$_x$TaS$_2$ as well as Ag$_x$TiS$_2$.~\cite{Scholz1980} There are further examples showing also superstructure effects in Ag$_x$NbS$_2$.~\cite{Wiegers1988,vanderLee1991}

In a recent publication we discussed the effects of alkali metal doping on the structural as well as electronic properties of \2 in addition to the changes in the band structure mentioned above~\cite{Brauer2001, Pettenkofer1992} and the crystal structure.~\cite{Omloo1970, Starnberg2000, Koenig2012} 
As was already shown in earlier reports, the CDW gets suppressed by doping,~\cite{Biberacher1985} accompanied by the formation of new potassium superstructures.~\cite{Koenig2012} In an associated theoretical investigation the structures of the various doping stages were modelled to estimate the potassium content of the \emph{in situ} doped samples. For the undoped case this band structure calculations show an agreement between the theoretical and experimental plasma frequency and the calculated doping rates therefore are used to estimate the experimental doping level throughout the present contribution.~\cite{Koenig2012}

In this report we present further results on the \emph{in situ} intercalation experiments of bulk samples of \2 with the alkali metal potassium. In particular, \eels\ (EELS) is applied to investigate changes in the plasmon dispersion. The possibility to measure the momentum dependence of electronic excitations is the particular strength of this method, as has been demonstrated previously in a number of cases.~\cite{Knupfer2000,Schuster2007,Kramberger2008,Roth2013} The investigations of the momentum dependence of the plasma resonance are accompanied by a \name{Kramers}-\name{Kronig} analysis (KKA) of our spectra.

\section{Experimental}
\label{sec:experimental}
 
Single crystals of \2 were grown from Ta metal and Se (\SI{99.95}{\percent} and \SI{99.999}{\percent} purity, respectively) by iodine vapor transport in a gradient of \SIrange{600}{620}{\degreeCelsius}, the crystals growing in the cooler end of the sealed quartz tubes. A very slight excess of Se was included (typically \SI{0.2}{\percent} of the charge) to ensure stoichiometry in the resulting crystals. Each experimental run lasted for \SIrange{250}{300}{h}. This procedure yielded single crystals with maximum size of $10 \times 10 \times \SI{0.2}{\cubic\milli\meter}$. Appropriate thin films (thickness of about \SI{100}{nm}) were prepared either by the use of an ultramicrotome or by repeatedly cleaving with adhesive tape. The films were mounted onto standard electron microscopy grids and transferred to the spectrometer. As already pointed out in Ref.\,\onlinecite{Koenig2012}, a very good sample quality is achieved, since clear superstructures of the CDW arising at the predicted transition temperature (\SI{120}{K}) are visible.

Potassium intercalation was achieved by evaporation from commercial SAES (SAES GETTERS S.\,P.\,A., Italy) sources under ultra high vacuum conditions (base pressure below \SI{e-10}{mbar}). The intercalation levels were achieved by several discrete intercalation steps, each of which took about \SI{3}{min}. The achieved doping concentrations were evaluated based on the procedure described in Ref.\,\onlinecite{Koenig2012}. Finally, the saturation potassium content was found to be stable over several months at room temperature.

All EELS measurements were carried out with a \SI{172}{kV} electron energy-loss spectrometer equipped with a He flow cryostat. The resolution of the spectrometer is \SI{0.03}{\per\angstrom} and \SI{80}{meV} for momentum and energy-loss, respectively. The spectrometer setup can be found in Ref.\,\onlinecite{Fink1989}. In order to carry out the KKA analysis, the quasi elastic line has been subtracted according to a procedure described previously. \cite{Schuster2009}

\section{Results and Discussion}
\label{sec:results}


\begin{figure*}
  \centering
  \includegraphics{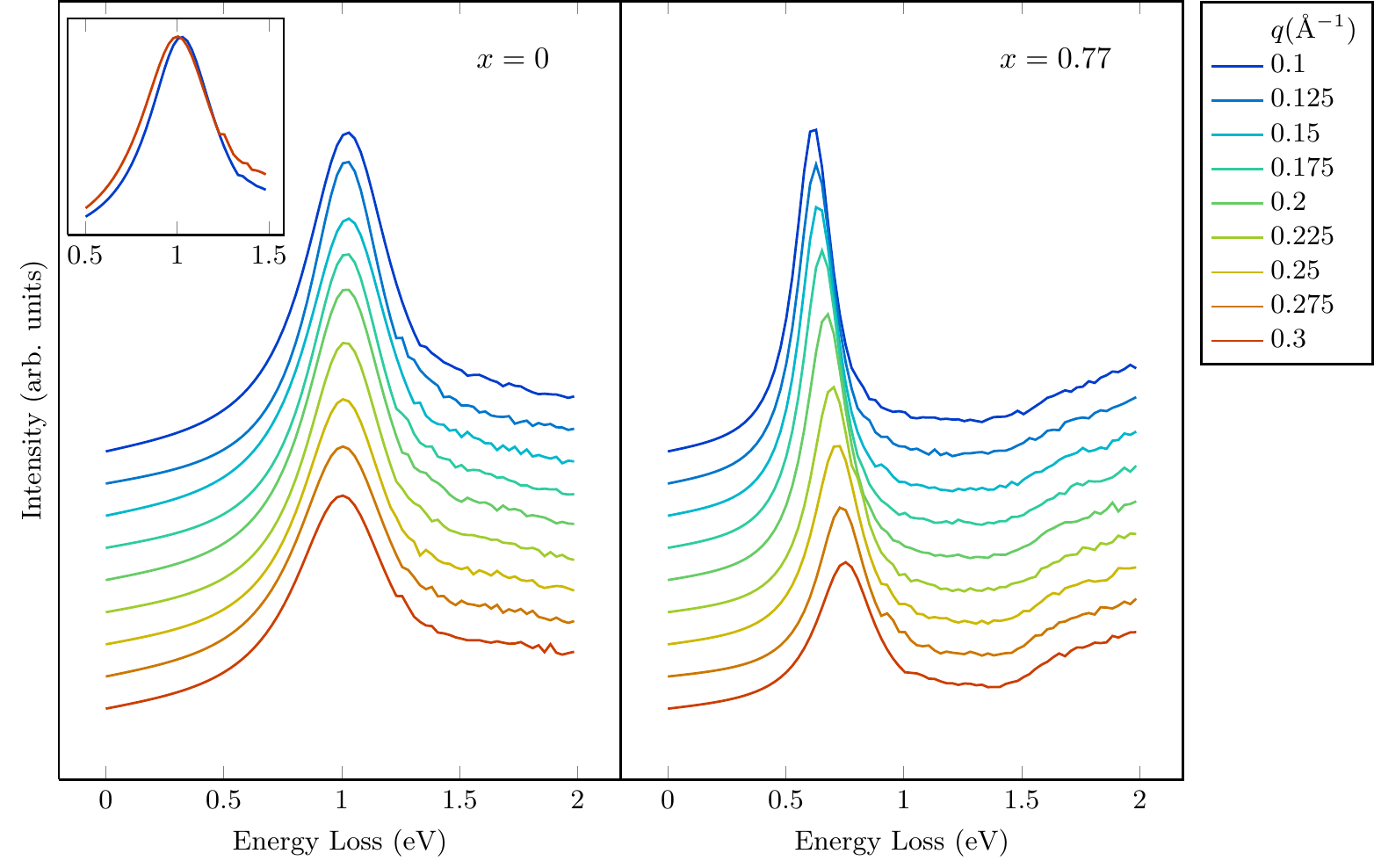}
  \caption{Energy loss spectra for zero as well as maximum doping for increasing momentum transfer values at room temperature. For illustration of the very small negative dispersion for $x=0$ the inset shows a direct comparison of the loss spectra for lowest and highest momentum transfer.}
  \label{fig:lossx}
\end{figure*}

The energy-loss spectra for three different doping stages 
are shown in \figref{fig:kka_spectra} in a wide energy range. Most prominently they show the volume plasmon at around \SI{21}{eV} splitting into a double peak structure on potassium addition. This splitting is a consequence of additional excitations from K\,3$p$ levels, which become possible at about \SI{18}{eV} for the potassium intercalated samples. The somewhat strange spectral shape of the loss function for the intercalated cases arises from the interaction of two energetically close lying (collective) excitations.~\cite{Widder1997} 
Further excitations as seen in \figref{fig:kka_spectra} are the well known \SI{1}{eV} charge carrier plasmon, shown in the inset and already discussed in Ref.\,\onlinecite{Koenig2012} and the higher lying excitations in the range of \SIrange{35}{60}{eV}. Primarily, the latter are excitations from the Ta\,5$p$ (\SI{42}{eV}) as well as the Se\,3$d$ (\SI{55}{eV}) core levels accompanied by multiple scattering effects of the volume plasmon.

The absolute value of the loss spectra were obtained applying a \name{Kramers}-\name{Kronig}-analysis (KKA). To this end, the measured spectra were extrapolated with an $\omega^{-3}$ dependence to achieve a broad range of integration, yielding the plotted loss functions.\cite{Livins1988} 
By fitting other functions derived from the KKA (e.\,g., the real part of the dielectric function $\epsilon_1$ as well as the optical conductivity $\sigma$) within the \name{Drude}-\name{Lorentz}~model,~\cite{Fink1989} the background dielectricity $\epsilon_{\infty}$ (being in the range of \SIrange{10}{15}{}) was found to be rather independent on the doping level of the sample.
Since this background dielectric constant describes the screening abilities of the interband transitions at energies above the plasmon energy, the observed energy shift of the charge-carrier plasmon on potassium intercalation as seen in the inset of \figref{fig:kka_spectra} predominantly results from the intercalation induced changes of the charge carriers. For a more detailed description of such an analysis of $\epsilon_{\infty}$ see Ref.\,\onlinecite{Roth2010}.

We take this as evidence that just the filling of the bands and not a change of their shape is the dominating intercalation effect within the basal plane of the crystal. We note however, that a rigid single band model,\cite{Campagnoli1979} where the doping is implemented by simply shifting the \name{Fermi} level to higher energies, is at odds with the multiband character of the \name{Fermi} surface.\cite{Borisenko2008}

Figure\,\ref{fig:lossx} shows energy-loss spectra in the energy regime around the charge carrier plasmon for the undoped as well as for the maximally doped ($x=0.77$) compound for increasing momentum transfer at room temperature. In the respective spectra for the lowest momentum transfer the plasmon peak shifts to lower energies when going from the undoped to the doped sample (as can be seen also from the inset of \figref{fig:lossx}). As discussed above, this is attributed to the varying charge carrier density, filling up the half filled band of the undoped compound.~\cite{Koenig2012, Campagnoli1979} Furthermore, the spectral width of the plasmon peak is smaller for the intercalated sample. Interband transitions that are damping the plasmon of the undoped sample might no longer be present in the lower energy regime of the shifted plasmon, which can provide a rationalization of the sharpening of the plasmon on doping.~\cite{Koenig2012}

Besides the slight broadening on increasing momentum transfer, it is obvious that the plasmon around \SI{1}{eV} exhibits a slight negative dispersion~\cite{Schuster2009} for the undoped compound (left panel of \figref{fig:lossx}). This behavior changes to a clear positive dependence for the fully doped material (right panel of \figref{fig:lossx}).

\begin{figure*}
  \centering
  \includegraphics{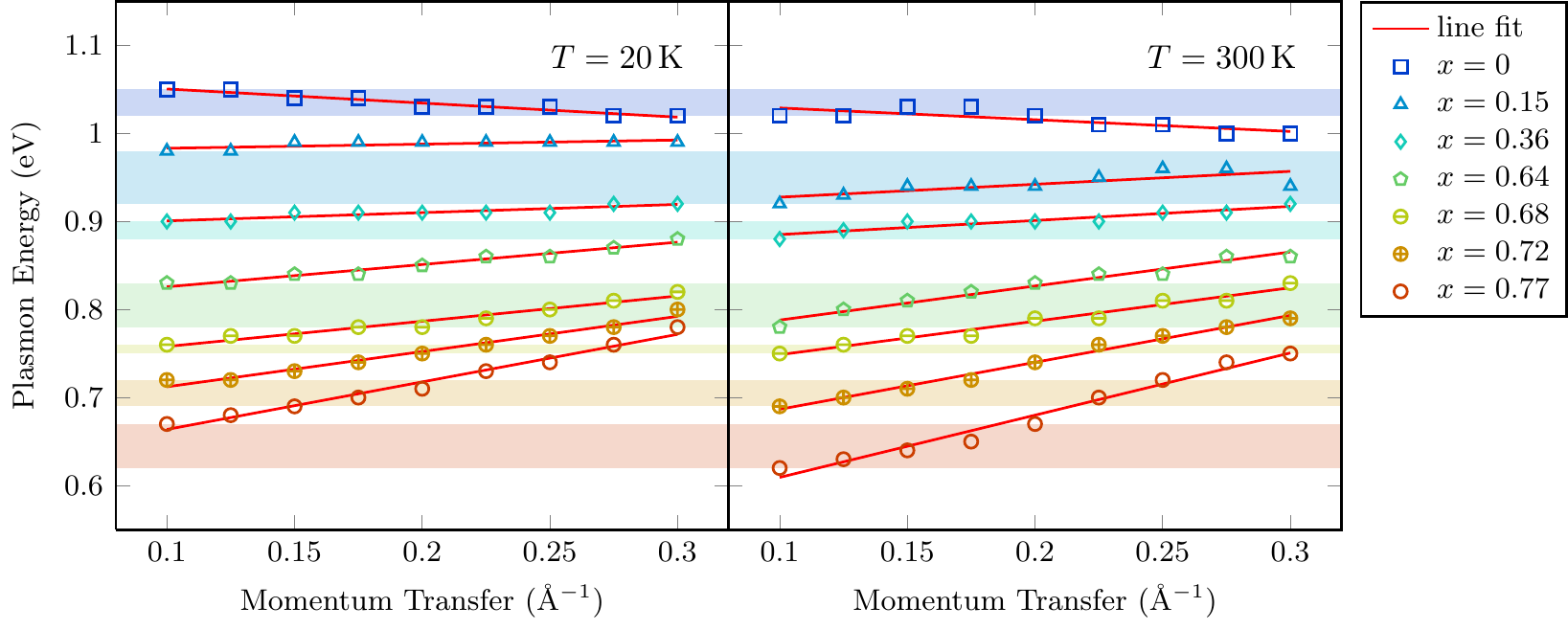}
  \caption{Dispersion of the plasmon for increasing potassium content and temperatures below and above the CDW phase transition. The red lines indicate the straight line fitting of the dispersions, while the colorbars represent the respective peak shifts of the plasmons at lowest momentum transfer.}
  \label{fig:dispx}
\end{figure*}

The evolution of the plasmon dispersion on potassium intercalation is evaluated in detail in \figref{fig:dispx}, where the dispersion of the \SI{1}{eV} plasmon as determined from \figref{fig:lossx} and additional measurements are shown. Being negative for the undoped sample the slope increases to values above zero on potassium addition. This behavior can be observed for low (\SI{20}{K}, left panel of \figref{fig:dispx}) as well as for high temperatures (\SI{300}{K}, right panel of \figref{fig:dispx}). The shaded bars indicate the blueshift of the plasmon energy going from higher to lower temperatures. This can be assigned to the enhanced charge density due to the contraction of the crystal structure on cooling. However, a doping dependence of this temperature induced shift can not be observed---it differs for all intercalation levels.

\begin{figure}
  \centering
  \includegraphics{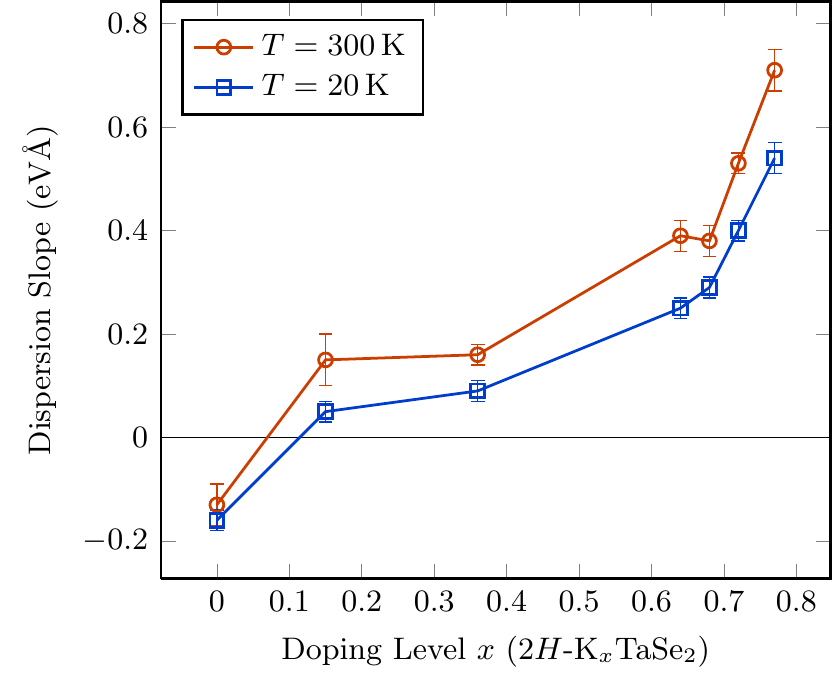}
  \caption{Slope of the plasmon dispersion of K$_x$TaSe$_2$ obtained by the linear fits shown in \figref{fig:dispx}.}
  \label{fig:disp_slope}
\end{figure}

To further analyze the changes in the dispersion in a more quantitative manner, the momentum dependencies plotted in \figref{fig:dispx} were fitted by straight lines (also shown in the figure). This might not describe the real momentum dependence of the dispersion~\cite{Schuster2009, vanWezel2011} but 
it provides a reasonable quantification of the observed variations at momentum transfers below \SI{0.3}{\per\angstrom}. The resulting slope values are plotted in \figref{fig:disp_slope}.

Notably, already for a small increase of the charge-carrier concentration by  \SI{15}{\percent} a substantial change in the dispersion is observed consisting of a large jump including a sign change. Besides a variation of the charge-carrier concentration, initial potassium addition also results in a larger crystal $c$~axis.~\cite{Koenig2012} In turn, the charge-carrier density also changes, and the associated downshift of the plasmon energy at smallest momentum transfer is, thus, a consequence of doping and lattice changes, while the latter is particularly prominent at initial K intercalation.~\cite{Koenig2012} The large jump in the plasmon dispersion, in particular the unusual change of sign, however, points towards deeper roots for this effect. It is important to realize that even at small doping of the order of \SI{15}{\percent} the charge density wave in $2H$-TMDC materials becomes considerably weakened if not suppressed.~\cite{Koenig2012,Shen2007,Fang2005} Moreover, the negative plasmon dispersion of undoped \2 has been rationalized in the framework of 
a semiclassical \name{Ginzburg}-\name{Landau} approach, and was assigned to an interplay of the charge density fluctuations and the plasma resonance.~\cite{vanWezel2011a} Keeping all this in mind, it is natural to ascribe the drastic changes in the plasmon dispersion for initial potassium addition in \2 to the suppression of the charge density wave in this compound on doping and the connected disappearance of the interaction of the charge density wave fluctuations with the charge-carrier plasmon. 

Intriguingly, adding more potassium (more than \SI{15}{\percent}) visibly affects the energy position of the plasmon at small momenta and the plasmon width (see figures above and Ref.\,\onlinecite{Koenig2012}), while the plasmon dispersion hardly changes until an intercalation of more than about \SI{60}{\percent} is reached. Above this value, there is a further significant rise of the slope of the plasmon dispersion. We also note that for all intercalation levels the charge-carrier plasmon is a well-defined spectral feature with decreasing width on increasing intercalation level (see also Ref.\,\onlinecite{Koenig2012}), which demonstrates that all measured potassium intercalated \2 compounds represent a homogeneous potassium distribution, i.\,e., the potassium intercalation occurs in a solid-solution manner. In view of a simple, free-electron gas description, one would expect that both the plasmon energy and the plasmon dispersion coefficient decrease with an effectively decreasing charge carrier concentration as induced by potassium addition.~\cite{Koenig2012,Campagnoli1979,Fink1989} Consequently, the significant rise of the plasmon dispersion as visualized in \figref{fig:disp_slope} represents a more complex phenomenon, and it might be connected to the multiband character of the \name{Fermi} surface of \2.\cite{Borisenko2008,Cudazzo2012,Faraggi2012}

Finally, the behavior on temperature variation as depicted in \figref{fig:dispx} and \figref{fig:disp_slope} parallels the observations as a function of potassium intercalation. While the larger plasmon energy at small momentum transfers for lower temperature is in line with the contraction of the lattice and the associated increase of the charge-carrier density, the slope of the plasmon dispersion is always smaller for lower temperature, which is opposite to what a simple model would predict. This issue needs further consideration and investigations in the future for final clarification.  

\section{Summary}
\label{sec:summary}

To summarize, we performed electron energy-loss spectroscopy studies of the transition-metal dichalcogenide \2 investigating the effect of intercalated potassium on the energy and dispersion of the charge-carrier plasmon. Analyzing 
the optical reponse functions of the gradually intercalated samples by a \name{Kramers}-\name{Kronig}~analysis, we argue that band filling is the dominating influence of the addition of the intercalate. The multiband character of the \name{Fermi} surface, however, contradicts a simple, rigid single band model. 

Our preparation procedure allowed rather high doping rates, and we demonstrate the momentum dependence of the charge-carrier plasmon to change in a very unusual manner from a negative to a positive slope with increasing potassium content. Remarkably, this behavior is observed for temperatures below the charge-density wave phase transition of the undoped compound as well as at room temperature.

A detailed analysis of this discontinuous change of the plasmon dispersion is achieved by fitting the dispersions by straight lines. 
While the shift of the plasmon energy at lowest momentum transfer is a consequence of the doping and the accompanying lattice change, the strong change in the dispersion at initial K intercalation is assigned to the suppression of the CDW, altering the coupling of the latter to the plasma resonance. The fact that the dispersion hardly changes with further doping  before a drastic change at high doping levels occurs might be caused by the multiband character of the compounds' \name{Fermi} surface. 

However, taking all this into account, it remains unclear why the closely related $2H$-NbS$_2$ is peculiar in some sense, since it is characterized by   a positive plasmon dispersion even for the undoped case.~\cite{Manzke1981} 
We suggest further investigations of this material class on Rb and Cs intercalation, in order to disentangle doping and lattice expansion effects. A comparison of the different $2H$-TMDCs as a function of intercalation might then also give deeper insight into the differences in regard to their plasmon dispersion and charge density formation.

\section*{Acknowledgement}

We thank \name{M.\,Naumann}, \name{S.\,Leger} and \name{R.\,Hübel} for technical assistance. This work was supported by the Deutsche Forschungsgemeinschaft via Grants No. KN393/12 and No. KN393/13.


%

\end{document}